\documentclass[reprint,superscriptaddress,showpacs,twocolumn,amsmath,amssymb,aps,prl]{revtex4}

\usepackage{graphicx}
\usepackage{dcolumn}
\usepackage{bm}
\usepackage{amssymb}
\usepackage{nicefrac}
\begin{document}

\author{Wolf B. Dapp}
\affiliation{J\"ulich Supercomputing Centre, Institute for Advanced Simulation, FZ J\"ulich, J\"ulich, Germany}
\author{Andreas L\"ucke}
\affiliation{J\"ulich Supercomputing Centre, Institute for Advanced Simulation, FZ J\"ulich, J\"ulich, Germany}
\affiliation{Department Physik, Universit\"at Paderborn, Paderborn, Germany}
\author{Bo N. J. Persson} 
\affiliation{Peter Gr\"unberg Institute, FZ J\"ulich, J\"ulich, Germany}
\author{Martin H. M\"user} 
\email{martin.mueser@mx.uni-saarland.de}
\affiliation{J\"ulich Supercomputing Centre, Institute for Advanced Simulation, FZ J\"ulich, J\"ulich, Germany}
\affiliation{Department of Materials Science and Engineering, Universit\"at des Saarlandes, Saarbr\"ucken, Germany}

\title{
Self-affine elastic contacts: percolation and leakage
}


\begin{abstract} 
We study fluid flow at the interfaces between elastic solids with randomly
rough, self-affine surfaces. We show by numerical simulation that elastic deformation
lowers the relative contact area at which contact patches percolate in comparison to
traditional approaches to seals. 
Elastic deformation also suppresses leakage through contacts even
far away from the percolation threshold.
Reliable estimates for leakage can be obtained by combining
Persson's contact mechanics theory with a slightly modified version
of Bruggeman's effective-medium solution of the Reynolds equation.
\end{abstract}

\maketitle

A seal is a device for closing a gap or making a joint fluid-tight~\cite{Flitney07}.
Although seals play a crucial role in many modern engineering devices,
inexpensive elastomeric seals such as O-rings are often used.
The failure of seals can have serious ramifications ranging from energy loss,
environmental pollution, expensive and time-consuming replacement 
procedures all the way to catastrophies like the Challenger disaster.
Thus, seal systems should be handled thoroughly in the design of machines,
and not like a secondary accessory.

Predicting leak rates is difficult, because the surface roughness
at the seal-substrate interface spans a wide range of length 
scales, from nanometers to centimeters~\cite{Persson05JPCM}.
For accurate leakage calculations, one first needs to identify the 
gap topography and then solve the Reynolds thin-film equation, in which
the local conductance is assumed to scale with the third power of the gap.
Despite significant progress in the recent past~\cite{bottiglione09, sahlin09,
sahlin09b, Persson05JPCM,Persson04JCP,Persson08JPCM,Lorenz09EPL,Lorenz10EPJE,
Persson10JPCM},
a comparison between large-scale numerical simulations (free of 
uncontrolled approximations) and approximate treatments is needed.

Industrial norms characterizing seal systems and traditional approaches to
derive the gap or gap distribution function use as input the cumulative 
height distribution function of the free, {\it undeformed} surfaces also known
as bearing area or Abbott and Firestone curve~\cite{Flitney07,Abbott33}. 
For example, gaps are constructed by simply ``cutting'' through the interface, 
i.e., $d(x,y) \equiv \max\{0, d_{\rm free}(x,y) - \Delta d\}$,
where $d_{\rm free}(x,y)$ is the gap for nontouching surfaces as 
a function of the lateral coordinates $x$ and $y$, and $\Delta d$
is a constant shift.

These ``bearing contacts'' disregard that material in the vicinity 
of a contact point is being pushed away from the interface (elastic deformation). 
However, neglecting elastic deformation induces
serious artifacts in contact mechanics. Relevant to seals are erroneous 
dependences of mean gap~\cite{Persson07PRL} and relative contact 
area~\cite{Carbone08} on normal load, as well as 
incorrect exponents for the contact autocorrelation 
functions~\cite{Ramisetti11}, indicating flawed contact
geometries. 
The contact geometry is crucial for the leakage problem,
because
stochastic indicators (such as the Euler characteristic~\cite{schmahling07}) 
in addition to the relative contact area $A/A_0$ (where $A$ is the true and 
$A_0$  the nominal contact area)
determine whether insulating contact patches or open channels 
percolate.
Therefore, approximations to the Reynolds thin-film equation that no 
longer contain information on the spatial arrangement of contact, such as 
Bruggeman's approach~\cite{Bruggeman35,Bo}, 
may jeopardize the results.

A promising approach to the leakage problem~\cite{Persson05JPCM,
Persson04JCP,Persson08JPCM,Lorenz10EPJE,Persson10JPCM} is based on the 
contact mechanics theory by Persson~\cite{Persson01}, which was 
developed for the friction between rubber and hard randomly
rough surfaces.
The starting point is the analysis of how  contact pressure 
or gap distribution functions broaden (on average for given surface 
height spectra) when finer and finer details of the surface topography are included 
in the calculation.
The approach reduces a high-dimensional partial
differential equation for the surface displacement to ordinary
differential equations for pressure or gap distribution functions.
Unlike traditional contact mechanics theories,
Persson theory produces correct functional dependencies
for the mean gap~\cite{Persson07PRL,Lorenz09JPCM,Akarapu11} and relative 
contact~\cite{Carbone08} on 
load as well as exact exponents for the contact 
autocorrelation function~\cite{Campana08,Persson08JPCMb}.
Lastly, Persson theory corresponds to a rigorous expansion
of an exact formulation of contact mechanics to {\it at least} third 
order in the inverse interfacial interaction range~\cite{Muser08PRL}.

When applied to leakage, Persson theory needs to be complemented with
approximate solvers to Reynolds' equation.
One approach is to use
Bruggeman's effective-medium theory~\cite{Bruggeman35}, which takes the 
gap distribution function 
as an input and predicts the percolation threshold to lie at a relative 
contact area of $A^*/A_0 = 1/2$.


Although Persson theory has predicted leakage and gap distribution functions in good agreement with
both experiment~\cite{Lorenz09EPL,Lorenz10EPJE,Lorenz10EPJE2} and numerical
approaches~\cite{Almqvist11}, some fundamental issues remain to be addressed.
First, the assumptions employed suffer to a certain degree from
uncontrollable uncertainties, i.e., the precision of roughness spectra of 
the free surfaces, role of shear thinning where flow gradients are large, 
and the exact slip boundary conditions.
The good agreement between theory and experiment may thus be 
partially fortuitous.
Conversely, cancellation of errors can be noticed in simulations
that realize all quantities to a defined precision.
Second, it is not clear where the percolation threshold is, how it 
is affected by elastic deformation, and how previous approaches need
to be altered,
should $A^*/A_0$ deviate from the ``canonical'' value of $1/2$~\cite{Zallen71}. 
Persson \textit{et al.} found unexpectedly small values for $A^*/A_0$ in
numerical simulations, but attributed this observation to finite-size 
effects in simulation cells of a linear size of ${\cal L} = 512$ grid 
points~\cite{Bo}. 
Other simulations~\cite{Carbone,Salin} also hint at the
possibility 
that elastic contacts may percolate below $A/A_0=1/2$.

In this Letter, we produce ``realistic'' gaps by solving the
elasticity equations for two rough solids in contact and solve the
Reynolds equation for the produced gaps without uncontrolled approximations.
We use ${\cal L} = 4096$, which is large enough to 
reflect the self-affinity of the surface topography and also ensure 
self-averaging of the fluid conductance.
This way we obtain percolation thresholds and leakage rates 
that are sufficiently accurate to determine the goodness of Bruggeman theory
in elastic contacts and if $A^*/A_0$ deviates
from $1/2$.

To solve the elastic problem, we use a slightly altered version of the 
Green's function molecular dynamics (GFMD) method presented in
Ref.~\cite{Campana06}.
First, we reduce the displacement field to a scalar,
thereby implicitly implementing the small-slope approximation~\cite{Johnson85}. 
Second, as Ref.~\cite{Akarapu11}, we use the continuum expression for
the elastic energy, i.e., 
$V_{\rm el} = \sum_{\bf q} E^* q \vert \tilde{z}({\bf q}) \vert^2/4$,
where ${\bf q}$ is an in-plane wave vector, $q$ its magnitude,
$E^*$ the effective elastic modulus, and $\tilde{z}({\bf q})$ the Fourier
transform of the normal displacement.
Third, we solve Newton's equations of motion in Fourier space but implement
the nonholonomic, hard-wall boundary conditions in real space.
Fourth, we damp the modes such that the slowest mode is critically
damped.
This way the relaxation time 
scales with $\sqrt{{\cal L}}$.
We map both compliance and roughness to
one side of the interface, as is allowed for our system~\cite{Johnson85}. 
The substrate topography is generated in Fourier space 
as described in Ref.~\cite{Campana08}; the height of the substrate
satisfies the rules for colored noise of self-affine fractals, i.e.,
$\langle \tilde{h}^*({\bf q}') \tilde{h}({\bf q}) \rangle \propto
\delta_{{\bf q}{\bf q}'} / q^{2+2H}$, where $H$ is the Hurst roughness
exponent. 
As a default, we allow for roughness between short and long
wavelengths cutoffs of $\lambda_{\rm s} = 1$, 
and $\lambda_{\rm l} = {\cal L}/8 = 512$, respectively, but vary
both bounds to reduce the risks of drawing false conclusions.

The gap topography produced in the GFMD simulation is used as the boundary 
condition for the Reynolds equation, which we solve with a 
central-differencing real-space method~\cite{numericalRecipes}. 
In order to speed up the calculations, we implemented a multi-grid 
preconditioner; i.e., we first solve the Reynolds equation on a coarse
grid, where the conductivity on each point is determined by invoking
Bruggeman theory on the subpoints.
The solution to the pressure on the coarse mesh is then interpolated onto
a finer grid on which the lattice constant is halved. 
This way, the initial guess for the fluid pressure is already 3 orders of
magnitude more accurate than the
mean field solution when we reach the finest resolution.

We first address percolation on {\it continuous} random 
domains~\cite{Zallen71}.
Given that the crossing of ``coastlines'' between 
contact and noncontact patches has zero measure in {\it two dimensions},
either contact or noncontact must percolate (except 
at the percolation threshold where stripes can occur). 
If the stochastic properties of contact at $A/A_0$ are identical to
those of noncontact at $1-A/A_0$ (as is the case for bearing contacts
of colored-noise surfaces),
the percolation threshold must lie at $A^*/A_0 = 1/2$.
We recover this value in our calculations, except for small scatter due to 
finite size.
Discretization effects are minor in our calculations, because the contact 
correlation length distinctly exceeds a lattice constant,
in particular for our default roughness exponent $H = 0.8$.
This differs from conventional lattice models where adjacent grid points 
are uncorrelated, which makes $A^*/A$ depend on the lattice
(simple cubic, hexagonal, etc.) and on the percolation type
(bond versus site percolation)~\cite{Stauffer91}. 

\begin{figure}[hbtp]
  \begin{minipage}[b]{0.5\columnwidth} 
    \centering
    \includegraphics[width=0.8\columnwidth]{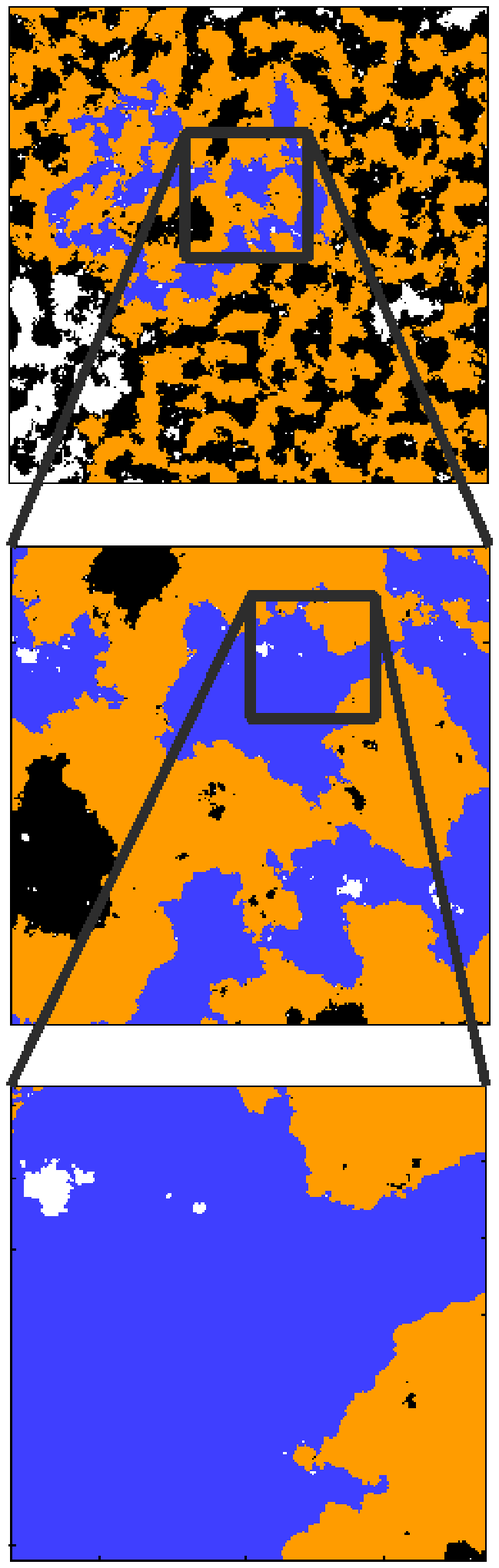}
  \end{minipage}%
  \hfill 
  \begin{minipage}[b]{0.5\columnwidth}
    \centering
    \includegraphics[width=0.8\columnwidth]{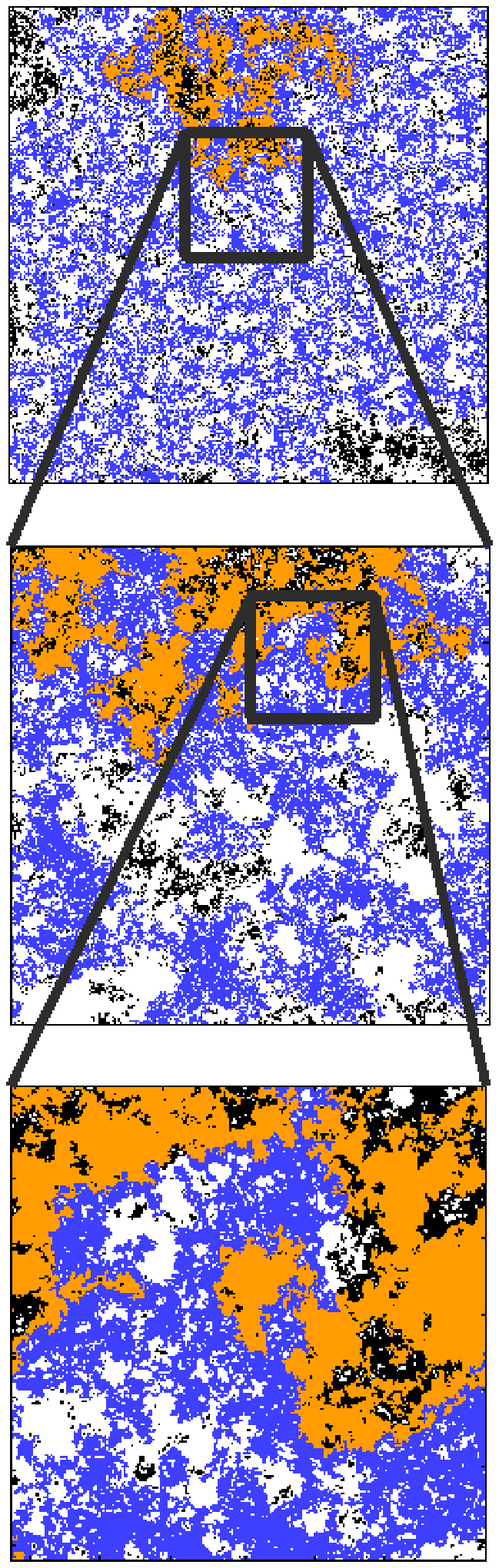}
  \end{minipage}
  \caption{ 
    Contact and noncontact patches for $A/A_0 = 0.46$
    and $H = 0.8$. 
    Black is regular contact, while dark gray (blue) represents the largest connected
    contact patch. 
    White and light gray (orange) represent similarly noncontact or open channels.
    Top panels show the full interface.
    \textbf{Left:} bearing-area model. \textbf{Right:} elastic calculations.}
  \label{fig:percolation}
\end{figure}

Including elastic deformation breaks the symmetry for the stochastic
properties of contact and noncontact patches, as one can see in
Fig.~\ref{fig:percolation}.
Noncontact now tends to break up into many small lubrication pockets, 
while the contact patches tend to form connected areas with holes
similar to ``Swiss cheese.'' 
In the language of algebraic topology, contact has a negative Euler 
characteristic and thus percolates more easily than noncontact
with a positive Euler characteristic~\cite{schmahling07}. 
Because of the symmetry-breaking of the stochastic properties for contact
and noncontact patches, elastic contacts have their percolation threshold
at $A^*/A_0 < 1/2$. 

As passing comments we note that contact is defined as zero gap between the 
two surfaces. We verified that any finite separation leads to vanishing forces 
between the GFMD layer and its counterface.
Furthermore, we find the same ratio of real contact area and load
as in continuum treatments~\cite{Hyun04,Campana07}, despite our choice
$\lambda_{\rm s} = 1$, because our elastic energy expression is that
of a continuous rather than a discrete system. 

Because of finite system size, a precise determination of $A^*/A_0$ 
remains difficult.
We estimate $A^*/A_0$ by taking the average value of 
$A/A_0$ where contact and noncontact start to percolate throughout the
system, respectively.
Owing to some remaining discretization effects, the width 
of the transition region where no ``color'' unambiguously 
dominates is $\Delta A /A_0 ({\cal L} = 4096) \approx \pm 0.02$. 
The parameters considered in our study encompass: 
$H = 0.4$ and $H = 0.8$, $1 \le \lambda_{\rm s} \le 4$, and
$512 \le \lambda_{\rm l} \le 2048$, as well as some disorder averaging 
over statistically equivalent surfaces (using different random seeds).
In all cases we find $A^*/A_0 = 0.5\pm 0.02$ for bearing contacts. 
This value is always reduced by $0.075\pm 0.015$ when the contact
is elastic, which leads us to an estimate of
$A^*/A_0 = 0.42(5)$. This is clearly less than the canonical value
of $1/2$ and at most weakly dependent on $H$. 

\begin{figure}[hbtp]
  \begin{minipage}[b]{0.5\columnwidth} 
    \centering
    \includegraphics[width=0.9\columnwidth]{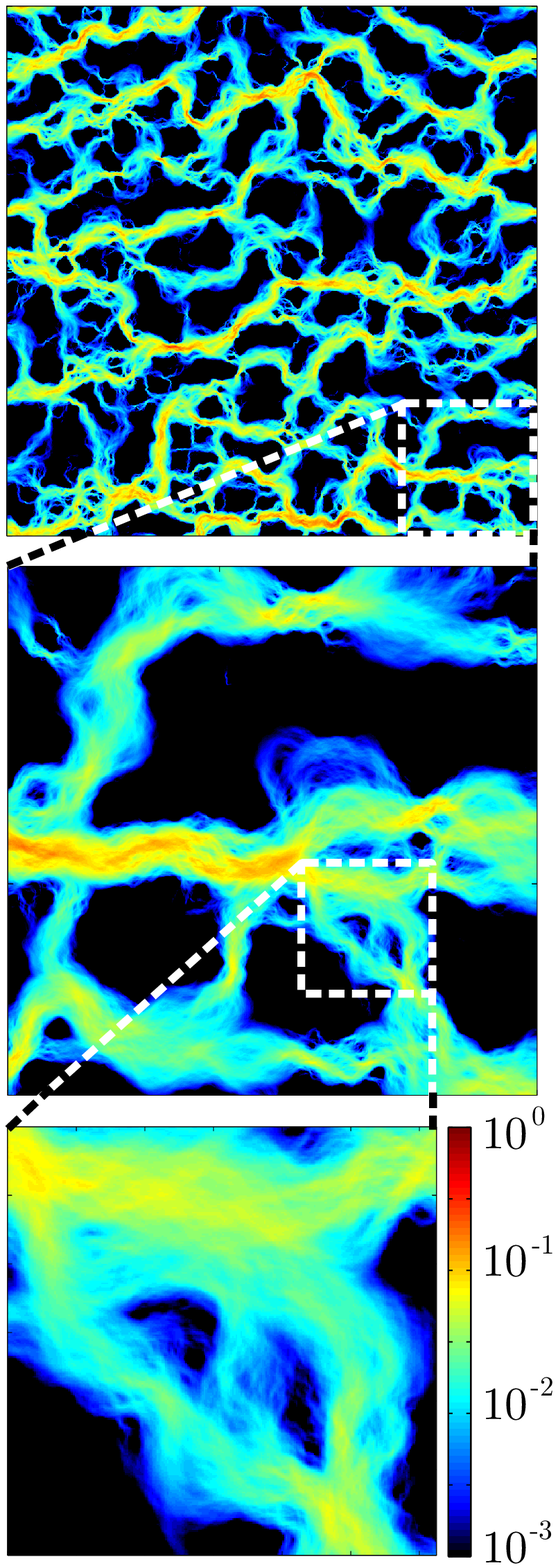}
  \end{minipage}%
  \hfill 
  \begin{minipage}[b]{0.5\columnwidth}
    \centering
    \includegraphics[width=0.9\columnwidth]{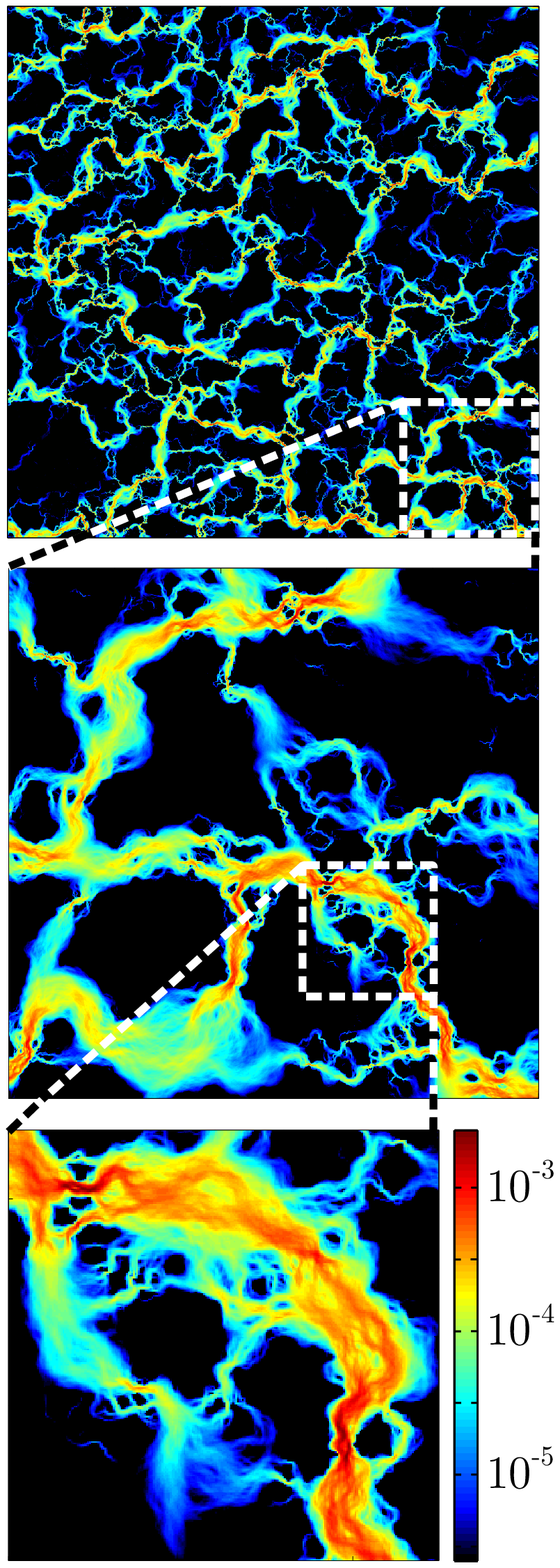}
  \end{minipage}
  \caption{
    Flow density through a contact with $A/A_0 = 0.2$.
    \textbf{Left:} bearing-area model. \textbf{Right:} elastic calculations.
    Note that the color scheme 
    between the two models differs by three decades. While generally the same
    channels are open, the flow is more constricted in the elastic case.
    Top panels show the full interface.} 
  \label{fig:leakage} 
\end{figure}

Since $A^*/A_0$ is smaller for elastic contacts one should expect reduced flow 
compared with bearing contacts. Indeed, the local maximum current intensities are
reduced by three decades.
Although the topography of the channel structure in elastic contacts resembles 
those of bearing contacts, Fig.~\ref{fig:leakage} shows the channels to be 
much narrower for elastic contact, even far away from percolation.

\begin{figure}[hbtp]
\includegraphics[width=1.0\columnwidth,angle=0]{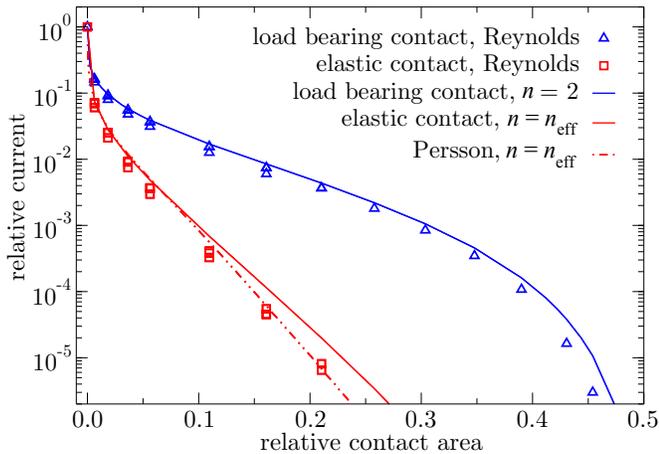}
\caption{
\label{fig:current}
Fluid current normalized by the value at first contact
as a function of relative contact area. 
Symbols represent data from numerical solutions of the Reynolds equation
for the bearing-area and elastic deformation models. 
Multiple symbols show surface topographies produced with different random 
seeds, and represent stochastic error.
Solid lines are predictions using the modified 
Bruggeman theory. The dot-dashed line is the prediction using Persson's 
contact mechanics theory and the modified Bruggeman theory.
}
\end{figure}

We produced both elastic and bearing contacts for a variety of loads.
Both types of contact topographies, more precisely their gap structures, 
were treated within the Bruggeman approximation~\cite{Bo}.
We obtained full solutions to Reynolds' equation only for a few loads, 
as those are very time consuming due to their slow convergence.
The results, shown in Fig.~\ref{fig:current}, demonstrate that current is 
strongly suppressed even far away from percolation. 
Furthermore, we find that the original Bruggeman theory is very accurate
for bearing contacts.
This is not surprising because Bruggeman theory is exact up to second
order in the gap fluctuation~\cite{Bo} and also produces the exact percolation
threshold for bearing contacts with colored-noise surface topographies.

For the elastic contact, we applied a small modification.
First, we note that the Bruggeman effective-medium theory in $n$-dimensional space predicts that
the noncontact area (for surfaces with roughness having isotropic statistical properties) 
percolates when $A^*/A_0 = (n-1)/n$, e.g., $A^*/A_0=1/2$
for $n=2$. In the self-consistent equation for the conductivity, we replaced the
physical dimension $n=2$ with an effective dimension $n_{\rm eff}(A)$.
For small contact area $A$, we want $n_{\rm eff}$ to be close to the physical dimension
of the interface, i.e., $n_{\rm eff}(0) = 2$, because Bruggeman is
essentially exact where $A\ll A_0$. 
However, in order to move the percolation to the correct location, we need
$n_{\rm eff} (A^*) = 1/(1-A^*/A_0)$, i.e., for $A^* = 0.42(5)~A_0$, 
$n_{\rm eff}(A^*) = 1.72$. 
In between these two extremes we interpolate linearly.
This is unnecessary for bearing-area surfaces, because they already 
have the Bruggeman threshold $A^*/A_0 = 1/2$.
While our modification lessens some of the beauty of the original approach,
Fig. \ref{fig:current} shows that it results in 
a very good agreement with the numerical solutions
over several decades in the conductivity. 
The even better agreement achieved with Persson theory is owed to
an $\mathcal{O}\left(10\%\right)$ underestimation of the gap,
which counteracts the overestimation of leakage in the
Bruggeman theory. 
 
To summarize, we have studied fluid flow at the interfaces between elastic solids with randomly
rough surfaces and show by numerical simulation that elastic deformation
lowers the relative contact area at which contact patches percolate 
[from $0.5$ to $\approx 0.42(5)$], and
suppresses leakage through such contacts even
far away from the percolation threshold, in comparison to
traditional approaches to seals. Leakage can be reliably estimated by combining
Persson's contact mechanics theory with a slightly modified version
of Bruggeman's effective-medium solution of the Reynolds equation.

\acknowledgments
We thank C. Denniston, N. Prodanov, and M.~O. Robbins for useful discussions
and the J\"ulich Supercomputing Centre for computing time.


\begin{thebibliography}{33}
\expandafter\ifx\csname natexlab\endcsname\relax\def\natexlab#1{#1}\fi
\expandafter\ifx\csname bibnamefont\endcsname\relax
  \def\bibnamefont#1{#1}\fi
\expandafter\ifx\csname bibfnamefont\endcsname\relax
  \def\bibfnamefont#1{#1}\fi
\expandafter\ifx\csname citenamefont\endcsname\relax
  \def\citenamefont#1{#1}\fi
\expandafter\ifx\csname url\endcsname\relax
  \def\url#1{\texttt{#1}}\fi
\expandafter\ifx\csname urlprefix\endcsname\relax\def\urlprefix{URL }\fi
\providecommand{\bibinfo}[2]{#2}
\providecommand{\eprint}[2][]{\url{#2}}

\bibitem[{\citenamefont{Flitney}(2007)}]{Flitney07}
\bibinfo{author}{\bibfnamefont{R.}~\bibnamefont{Flitney}},
  \emph{\bibinfo{title}{Seals and Sealing handbook}}
  (\bibinfo{publisher}{Elsevier}, \bibinfo{address}{New York}, \bibinfo{year}{2007}).

\bibitem[{\citenamefont{{B.~N.~J. Persson {\it et al.}}}(2005)}]{Persson05JPCM}
\bibinfo{author}{\bibnamefont{{B.~N.~J. Persson, O. Albohr, U. Tartaglino, A. I. Volokitin,
and E. Tosatti}}},
  \bibinfo{journal}{J. Phys. Condens. Matter} \textbf{\bibinfo{volume}{17}},
  \bibinfo{pages}{R1} (\bibinfo{year}{2005}).

\bibitem[{\citenamefont{{F. Bottiglione {\it et al.}}}(2009)}]{bottiglione09}
\bibinfo{author}{\bibnamefont{{F. Bottiglione, G. Carbone, L. Mangialardi, and G.
Mantriota}}},
  \bibinfo{journal}{J. Appl. Phys.} \textbf{\bibinfo{volume}{106}},
  \bibinfo{pages}{104902} (\bibinfo{year}{2009}).

\bibitem[{\citenamefont{Sahlin et~al.}(2009{\natexlab{a}})\citenamefont{Sahlin,
  Larsson, Almqvist, Lugt, and Marklund}}]{sahlin09}
\bibinfo{author}{\bibfnamefont{F.}~\bibnamefont{Sahlin}},
  \bibinfo{author}{\bibfnamefont{R.}~\bibnamefont{Larsson}},
  \bibinfo{author}{\bibfnamefont{A.}~\bibnamefont{Almqvist}},
  \bibinfo{author}{\bibfnamefont{P.~M.} \bibnamefont{Lugt}}, \bibnamefont{and}
  \bibinfo{author}{\bibfnamefont{P.}~\bibnamefont{Marklund}},
  \bibinfo{journal}{Proc. Instn. Mech. Engrs. Part J} \textbf{\bibinfo{volume}{224}},
  \bibinfo{pages}{335} (\bibinfo{year}{2009}{\natexlab{a}}).

\bibitem[{\citenamefont{Sahlin et~al.}(2009{\natexlab{b}})\citenamefont{Sahlin,
  Larsson, Marklund, Almqvist, and Lugt}}]{sahlin09b}
\bibinfo{author}{\bibfnamefont{F.}~\bibnamefont{Sahlin}},
  \bibinfo{author}{\bibfnamefont{R.}~\bibnamefont{Larsson}},
  \bibinfo{author}{\bibfnamefont{P.}~\bibnamefont{Marklund}},
  \bibinfo{author}{\bibfnamefont{A.}~\bibnamefont{Almqvist}}, \bibnamefont{and}
  \bibinfo{author}{\bibfnamefont{P.~M.} \bibnamefont{Lugt}},
  \bibinfo{journal}{Proc. Instn. Mech. Engrs. Part J} \textbf{\bibinfo{volume}{224}},
  \bibinfo{pages}{353} (\bibinfo{year}{2009}{\natexlab{b}}).

\bibitem[{\citenamefont{Persson et~al.}(2004)\citenamefont{Persson, Albohr,
  Creton, and Peveri}}]{Persson04JCP}
\bibinfo{author}{\bibfnamefont{B.~N.~J.} \bibnamefont{Persson}},
  \bibinfo{author}{\bibfnamefont{O.}~\bibnamefont{Albohr}},
  \bibinfo{author}{\bibfnamefont{C.}~\bibnamefont{Creton}}, \bibnamefont{and}
  \bibinfo{author}{\bibfnamefont{V.}~\bibnamefont{Peveri}},
  \bibinfo{journal}{J. Chem. Phys.} \textbf{\bibinfo{volume}{120}},
  \bibinfo{pages}{8779} (\bibinfo{year}{2004}).

\bibitem[{\citenamefont{Persson and Yang}(2008)}]{Persson08JPCM}
\bibinfo{author}{\bibfnamefont{B.~N.~J.} \bibnamefont{Persson}}
  \bibnamefont{and} \bibinfo{author}{\bibfnamefont{C.}~\bibnamefont{Yang}},
  \bibinfo{journal}{J. Phys. Condens. Matter} \textbf{\bibinfo{volume}{20}},
  \bibinfo{pages}{315011} (\bibinfo{year}{2008}).

\bibitem[{\citenamefont{Lorenz and Persson}(2009{\natexlab{a}})}]{Lorenz09EPL}
\bibinfo{author}{\bibfnamefont{B.}~\bibnamefont{Lorenz}} \bibnamefont{and}
  \bibinfo{author}{\bibfnamefont{B.~N.~J.} \bibnamefont{Persson}},
  \bibinfo{journal}{EPL} \textbf{\bibinfo{volume}{86}}, \bibinfo{pages}{44006}
  (\bibinfo{year}{2009}{\natexlab{a}}).

\bibitem[{\citenamefont{Lorenz and Persson}(2010{\natexlab{a}})}]{Lorenz10EPJE}
\bibinfo{author}{\bibfnamefont{B.}~\bibnamefont{Lorenz}} \bibnamefont{and}
  \bibinfo{author}{\bibfnamefont{B.~N.~J.} \bibnamefont{Persson}},
  \bibinfo{journal}{Eur. Phys. J. E} \textbf{\bibinfo{volume}{31}},
  \bibinfo{pages}{159} (\bibinfo{year}{2010}{\natexlab{a}}).

\bibitem[{\citenamefont{Persson}(2010)}]{Persson10JPCM}
\bibinfo{author}{\bibfnamefont{B.~N.~J.} \bibnamefont{Persson}},
  \bibinfo{journal}{J. Phys. Condens. Matter} \textbf{\bibinfo{volume}{22}},
  \bibinfo{pages}{265004} (\bibinfo{year}{2010}).

\bibitem[{\citenamefont{Abbott and Firestone}(1933)}]{Abbott33}
\bibinfo{author}{\bibfnamefont{E.~J.} \bibnamefont{Abbott}} \bibnamefont{and}
  \bibinfo{author}{\bibfnamefont{F.~A.} \bibnamefont{Firestone}},
  \bibinfo{journal}{Mech. Eng.} \textbf{\bibinfo{volume}{55}},
  \bibinfo{pages}{569} (\bibinfo{year}{1933}).

\bibitem[{\citenamefont{Persson}(2007)}]{Persson07PRL}
\bibinfo{author}{\bibfnamefont{B.~N.~J.} \bibnamefont{Persson}},
  \bibinfo{journal}{Phys. Rev. Lett.} \textbf{\bibinfo{volume}{99}},
  \bibinfo{pages}{125502} (\bibinfo{year}{2007}).

\bibitem[{\citenamefont{Carbone and Bottiglione}(2008)}]{Carbone08}
\bibinfo{author}{\bibfnamefont{G.}~\bibnamefont{Carbone}} \bibnamefont{and}
  \bibinfo{author}{\bibfnamefont{F.}~\bibnamefont{Bottiglione}},
  \bibinfo{journal}{J. Mech. Phys. Solids} \textbf{\bibinfo{volume}{56}},
  \bibinfo{pages}{2555} (\bibinfo{year}{2008}).

\bibitem[{\citenamefont{{S.~B. Ramisetti {\it et al.}}}(2011)}]{Ramisetti11}
\bibinfo{author}{\bibnamefont{{S.~B. Ramisetti {\it et al.}}}},
  \bibinfo{journal}{J. Phys.: Condens. Matter} \textbf{\bibinfo{volume}{23}},
  \bibinfo{pages}{215004} (\bibinfo{year}{2011}).

\bibitem[{\citenamefont{Schm\"ahling and Hamprecht}(2007)}]{schmahling07}
\bibinfo{author}{\bibfnamefont{J.}~\bibnamefont{Schm\"ahling}}
  \bibnamefont{and} \bibinfo{author}{\bibfnamefont{F.~A.}
  \bibnamefont{Hamprecht}}, \bibinfo{journal}{Wear}
  \textbf{\bibinfo{volume}{262}}, \bibinfo{pages}{1360} (\bibinfo{year}{2007}).

\bibitem[{\citenamefont{Bruggeman}(1935)}]{Bruggeman35}
\bibinfo{author}{\bibfnamefont{D.~A.~G.} \bibnamefont{Bruggeman}},
  \bibinfo{journal}{Ann. Phys. Lpz.} \textbf{\bibinfo{volume}{24}},
  \bibinfo{pages}{636} (\bibinfo{year}{1935}).

\bibitem[{\citenamefont{{B.~N.~J. Persson {\it et al.}}}(2012)}]{Bo}
\bibinfo{author}{\bibnamefont{{B.~N.~J. Persson {\it et al.}}}},
  \bibinfo{journal}{Eur. Phys. J. E} \textbf{\bibinfo{volume}{35}},
  \bibinfo{pages}{5} (\bibinfo{year}{2012}).

\bibitem[{\citenamefont{Persson}(2001)}]{Persson01}
\bibinfo{author}{\bibfnamefont{B.~N.~J.} \bibnamefont{Persson}},
  \bibinfo{journal}{J. Chem. Phys.} \textbf{\bibinfo{volume}{115}},
  \bibinfo{pages}{3840} (\bibinfo{year}{2001}).

\bibitem[{\citenamefont{Lorenz and Persson}(2009{\natexlab{b}})}]{Lorenz09JPCM}
\bibinfo{author}{\bibfnamefont{B.}~\bibnamefont{Lorenz}} \bibnamefont{and}
  \bibinfo{author}{\bibfnamefont{B.~N.~J.} \bibnamefont{Persson}},
  \bibinfo{journal}{J. Phys. Condens. Matter} \textbf{\bibinfo{volume}{21}},
  \bibinfo{pages}{015003} (\bibinfo{year}{2009}{\natexlab{b}}).

\bibitem[{\citenamefont{Akarapu et~al.}(2011)\citenamefont{Akarapu, Sharp, and
  Robbins}}]{Akarapu11}
\bibinfo{author}{\bibfnamefont{S.}~\bibnamefont{Akarapu}},
  \bibinfo{author}{\bibfnamefont{T.}~\bibnamefont{Sharp}}, \bibnamefont{and}
  \bibinfo{author}{\bibfnamefont{M.~O.} \bibnamefont{Robbins}},
  \bibinfo{journal}{Phys. Rev. Lett.} \textbf{\bibinfo{volume}{106}},
  \bibinfo{pages}{204301} (\bibinfo{year}{2011}).

\bibitem[{\citenamefont{Campa{\~n}\'a et~al.}(2008)\citenamefont{Campa{\~n}\'a,
  M\"user, and Robbins}}]{Campana08}
\bibinfo{author}{\bibfnamefont{C.}~\bibnamefont{Campa{\~n}\'a}},
  \bibinfo{author}{\bibfnamefont{M.~H.} \bibnamefont{M\"user}},
  \bibnamefont{and} \bibinfo{author}{\bibfnamefont{M.~O.}
  \bibnamefont{Robbins}}, \bibinfo{journal}{J. Phys.: Condens. Matter}
  \textbf{\bibinfo{volume}{20}}, \bibinfo{pages}{354013}
  (\bibinfo{year}{2008}).

\bibitem[{\citenamefont{Persson}(2008)}]{Persson08JPCMb}
\bibinfo{author}{\bibfnamefont{B.~N.~J.} \bibnamefont{Persson}},
  \bibinfo{journal}{J. Phys. Condens. Matter} \textbf{\bibinfo{volume}{20}},
  \bibinfo{pages}{312001} (\bibinfo{year}{2008}).

\bibitem[{\citenamefont{M\"user}(2008)}]{Muser08PRL}
\bibinfo{author}{\bibfnamefont{M.~H.} \bibnamefont{M\"user}},
  \bibinfo{journal}{Phys. Rev. Lett.} \textbf{\bibinfo{volume}{100}},
  \bibinfo{pages}{055504} (\bibinfo{year}{2008}).

\bibitem[{\citenamefont{Lorenz and
  Persson}(2010{\natexlab{b}})}]{Lorenz10EPJE2}
\bibinfo{author}{\bibfnamefont{B.}~\bibnamefont{Lorenz}} \bibnamefont{and}
  \bibinfo{author}{\bibfnamefont{B.~N.~J.} \bibnamefont{Persson}},
  \bibinfo{journal}{Eur. Phys. J. E} \textbf{\bibinfo{volume}{32}},
  \bibinfo{pages}{281} (\bibinfo{year}{2010}{\natexlab{b}}).

\bibitem[{\citenamefont{Almqvist et~al.}(2011)\citenamefont{Almqvist,
  Campa{\~n}\'a, Prodanov, and Persson}}]{Almqvist11}
\bibinfo{author}{\bibfnamefont{A.}~\bibnamefont{Almqvist}},
  \bibinfo{author}{\bibfnamefont{C.}~\bibnamefont{Campa{\~n}\'a}},
  \bibinfo{author}{\bibfnamefont{N.}~\bibnamefont{Prodanov}}, \bibnamefont{and}
  \bibinfo{author}{\bibfnamefont{B.~N.~J.} \bibnamefont{Persson}},
  \bibinfo{journal}{J. Mech. Phys. Solids} \textbf{\bibinfo{volume}{59}},
  \bibinfo{pages}{2355} (\bibinfo{year}{2011}).

\bibitem[{\citenamefont{Zallen and Scher}(1971)\citenamefont{Zallen and Scher}}]{Zallen71}
\bibinfo{author}{\bibfnamefont{R.}~\bibnamefont{Zallen}}
  \bibnamefont{and}
  \bibinfo{author}{\bibfnamefont{H.}~\bibnamefont{Scher}},
  \bibinfo{journal}{Phys. Rev. B} \textbf{\bibinfo{volume}{4}},
  \bibinfo{pages}{4471} (\bibinfo{year}{1971}).

\bibitem[{Car()}]{Carbone}
\bibinfo{note}{G. Carbone, private communication}.

\bibitem[{\citenamefont{Sahlin}(2008)}]{Salin}
\bibinfo{author}{\bibfnamefont{F.}~\bibnamefont{Sahlin}}, Ph.D. thesis,
  \bibinfo{school}{Lule{\aa} University of Technology} (\bibinfo{year}{2008});
  \bibinfo{note}{see, in particular, appendix F}.

\bibitem[{\citenamefont{Campa{\~n}\'a and M\"user}(2006)}]{Campana06}
\bibinfo{author}{\bibfnamefont{C.}~\bibnamefont{Campa{\~n}\'a}}
  \bibnamefont{and} \bibinfo{author}{\bibfnamefont{M.~H.}
  \bibnamefont{M\"user}}, \bibinfo{journal}{Phys. Rev. B}
  \textbf{\bibinfo{volume}{74}}, \bibinfo{pages}{075420}
  (\bibinfo{year}{2006}).

\bibitem[{\citenamefont{Johnson}(1985)}]{Johnson85}
\bibinfo{author}{\bibfnamefont{K.~L.} \bibnamefont{Johnson}},
  \emph{\bibinfo{title}{Contact Mechanics}} (\bibinfo{publisher}{Cambridge
  University Press}, \bibinfo{year}{1985}).

\bibitem[{\citenamefont{{\it et al.}}(2007)}]{numericalRecipes}
\bibinfo{author}{\bibfnamefont{W.~H.~Press} \bibnamefont{{\it et al.}}},
  \emph{\bibinfo{title}{Numerical Recipes}} (\bibinfo{publisher}{Cambridge
  University Press}, \bibinfo{address}{Cambridge, England}, \bibinfo{year}{2007}).


\bibitem[{\citenamefont{Stauffer and Aharony}(1991)}]{Stauffer91}
\bibinfo{author}{\bibfnamefont{D.}~\bibnamefont{Stauffer}} \bibnamefont{and}
  \bibinfo{author}{\bibfnamefont{A.}~\bibnamefont{Aharony}},
  \emph{\bibinfo{title}{An Introduction to Percolation Theory}}
  (\bibinfo{publisher}{CRC}, \bibinfo{year}{1991}).

\bibitem[{\citenamefont{Hyun et~al.}(2004)\citenamefont{Hyun, Pei, Molinari,
  and Robbins}}]{Hyun04}
\bibinfo{author}{\bibfnamefont{S.}~\bibnamefont{Hyun}},
  \bibinfo{author}{\bibfnamefont{L.}~\bibnamefont{Pei}},
  \bibinfo{author}{\bibfnamefont{J.-F.} \bibnamefont{Molinari}},
  \bibnamefont{and} \bibinfo{author}{\bibfnamefont{M.~O.}
  \bibnamefont{Robbins}}, \bibinfo{journal}{Phys. Rev. E}
  \textbf{\bibinfo{volume}{70}}, \bibinfo{pages}{026117}
  (\bibinfo{year}{2004}).

\bibitem[{\citenamefont{Campa{\~n}\'a and M\"user}(2007)}]{Campana07}
\bibinfo{author}{\bibfnamefont{C.}~\bibnamefont{Campa{\~n}\'a}}
  \bibnamefont{and} \bibinfo{author}{\bibfnamefont{M.~H.}
  \bibnamefont{M\"user}}, \bibinfo{journal}{EPL} \textbf{\bibinfo{volume}{77}},
  \bibinfo{pages}{38005} (\bibinfo{year}{2007}).

\end{thebibliography}

\end{document}